\documentclass{appolb}
\usepackage{epsfig}
\newcommand{\eep}{\mbox{(e,e$'$p)}}
\def\bra#1{\left\langle #1\right|}
\def\ket#1{\left| #1\right\rangle}
\begin{document}
\title{Nuclear Equation of State and Spectral Functions 
\thanks{Presented at XXVII Mazurian Lakes School of Physics, ``Growth Points
of Nuclear Physics A.D. 2001'', Krzyze, Poland, September 3-8, 2001}%
}
\author{W. H. Dickhoff
\address{Laboratory of Theoretical Physics, University of Gent, \\
Proeftuinstraat 86, B-9000 Gent, Belgium \\
Department of Physics, Washington University, St. Louis, Missouri 63130, USA}
\and
E. P.Roth
\address{Department of Physics, Washington University,
St. Louis, Missouri 63130, USA}
}
\maketitle
\begin{abstract}
  An overview is given of the theoretical work on
  nucleon spectral functions in finite nuclei. 
  The consequences of the observed spectral strength distribution
 are then considered in the context of the nuclear-matter saturation 
problem. Arguments are presented suggesting that
  short-range correlations are mainly responsible for the actual value of 
  the observed charge density in ${}^{208}\rm Pb$ and by extension for the 
  empirical value of the saturation density of nuclear matter.  
  This observation combined with the general understanding of the spectroscopic
  strength suggests that a renewed study of nuclear matter, emphasizing the
  self-consistent determination of the spectral strength due to short-range
  and tensor correlations, may shed light on the perennial
  nuclear saturation problem.
  First results using such a scheme are presented.
  Arguments are discussed that clarify the role of long-range correlations and 
their relevance for nuclear saturation.
\end{abstract}
  
\section{Introduction}
During the last fifteen years considerable progress has been made in 
clarifying the limits of the nuclear mean-field picture.
The primary tool in exhibiting these limits in a quantitative fashion
has been provided by the \eep\ reaction [1-4].
In this paper the status of the theoretical understanding of the spectroscopic 
factors that have been deduced from the analysis of this reaction will be
briefly reviewed.
The qualitative features of the strength distribution can be understood by 
realizing that a considerable mixing occurs between hole states and two-hole 
one-particle (2h1p) states. This leads to the observed fragmentation pattern
which exhibits a single peak for valence hole states near the Fermi energy,
albeit with a reduction of the strength by about 35\% [1-4].
A broadly fragmented strength distribution 
is observed for more deeply bound states which also sums to about
65\% of the strength.
This strong fragmentation is due
to the strong coupling of these single-hole states to 2h1p states and the 
accompanying small energy denominators.
For quantitative results one also 
requires the inclusion of short-range and tensor
correlations. 
First, this leads to a global depletion of mean-field
orbitals which ranges from 10\% in light nuclei to about 15\% in heavy nuclei
and nuclear matter [5,6].
Second, this depletion effect must be
compensated by the admixture of high-momentum
components in the ground state.
Such high-momentum nucleons have not yet been unambiguously identified 
experimentally using the \eep\ reaction. The search for such high-momentum
components in valence hole states has not been successful [7,8] as
was predicted by preceding theoretical work [9]. 
More details on some of these issues will be discussed in Sect.~2.

A recent publication [10] has challenged the conventional
interpretation of the \eep\ reaction with regard to valence hole states.
This challenge consists in questioning the validity of the constancy of the 
spectroscopic factor as a function of the four-momentum $Q^2$, transferred by 
the virtual photon to the knocked-out nucleon.
While the theoretical definition of the spectroscopic factor is unambiguously
independent of the probe, it is worth studying the description of the data at 
higher $Q^2$ in a consistent manner.
In Ref.~[11] this approach is outlined for the description of a
recent JLab experiment [12]. 
An eikonal description of the final-state-interaction [13] was combined with
previous results for the quasihole wave functions obtained for
${}^{16}$O [14] which were employed for the description [15]
of a low $Q^2$ experiment performed at NIKHEF [16]. 
The absorption of the outgoing proton was
related to the absorption experienced by a nucleon with corresponding momentum
in nuclear matter.
This information was obtained from self-consistent calculations for 
nucleon spectral functions discussed in more detail below [17].
The results at $Q^2 = 0.8 {\rm (GeV/c)^2}$ 
[11] demonstrate that the same spectroscopic factors can
be employed for a successful description of the JLab experiment [12] 
{\bf and} the earlier NIKHEF data [16] confirming
the usual interpetation of the \eep\ reaction. 

The information about the spectral strength
distribution provides new motivation to consider the ``energy'' or
``Koltun'' sum rule [18,19].
In principle, one can observe that a perfect agreement of the theoretical
strength with the experimental one, at all energies and all momenta,
must yield a correspondingly good agreement
for the energy per particle, provided three-body forces are not too important.
The importance of the contribution of high-momentum nucleons, which sofar
have not been observed directly, to the energy per 
particle has already been pointed out in [14].
In addition, we will argue in Sect.~3 that the actual value of the
nuclear saturation density is dominated by the effects of short-range (and 
tensor) correlations (SRC).
Recent experimental work supports this claim.
Based on these considerations, it is argued that a renewed study of the
nuclear satuation problem is in order.
Special emphasis on SRC will be utilized and all possible
contributions of long-range correlations discarded.
First results of this computationally demanding scheme 
will be discussed in some detail in Sect.~4.
Finally, some conclusions are presented in Sect.~5.

\section{Status of Theoretical Results for Spectroscopic Strength}
Exclusive experiments, involving the removal of a proton from the nucleus
which is induced by a high-energy electron that
is detected in coincidence with the
removed proton, have given access to absolute spectroscopic factors
associated with quasihole states for a wide range of
nuclei [1-4].
The experimental results indicate that the removal of single-particle (sp)
strength for quasihole states
near the Fermi energy corresponds to about 65\%.
The spectroscopic factors obtained in these experiments can be directly related
to the sp Green's function of the system which is given by
\begin{eqnarray}
G(\alpha , \beta ; \omega ) & = &
\sum_m\ \frac{\bra{\Psi_0^A} a_\alpha \ket{\Psi_m^{A+1}}
\bra{\Psi_m^{A+1}} a_\beta^\dagger \ket{\Psi_0^A}}
{\omega - (E_m^{A+1} - E_0^A ) + i\eta} \nonumber \\
& + & \sum_n\ \frac{\bra{\Psi_0^A} a_\beta^\dagger \ket{\Psi_n^{A-1}}
\bra{\Psi_n^{A-1}} a_\alpha \ket{\Psi_0^A}}
{\omega - (E_0^A - E_n^{A-1} ) - i\eta} .
\label{eq:prop}
\end{eqnarray}
This representation of the Green's function is referred to as the 
Lehmann-representation and involves the exact eigenstates 
and corresponding energies of the
$A$- and $A\pm 1$-particle systems.
Both the addition and removal amplitude for a particle from (to)
the ground state of the system with $A$ particles must be considered
in Eq.~(\ref{eq:prop}).
Only the removal amplitude has direct relevance for the analysis
of the \eep\ experiments.
The spectroscopic factor for the removal of a particle in the sp
orbit $\alpha$, while leaving the remaining nucleus in state $n$,
is then given by
\begin{equation}
z_\alpha^n = \left| \bra{\Psi_n^{A-1}} a_\alpha \ket{\Psi_0^A} \right|^2 ,
\label{eq:specf}
\end{equation}
which corresponds to the contribution to the numerator of the second sum
in Eq.~(\ref{eq:prop}) of state $n$ for the case $\beta = \alpha$.
Another important quantity is the spectral
function associated with sp orbit $\alpha$.
The part related to the removal of particles, or hole spectral function,
is given by
\begin{equation}
S_h(\alpha , \omega) = \sum_n\
\left| \bra{\Psi_n^{A-1}} a_\alpha \ket{\Psi_0^A} \right|^2 
\delta(\omega - (E_0^A - E_n^{A-1} ) ) ,
\label{eq:spefu}
\end{equation}
which corresponds to the imaginary part of the diagonal elements
of the propagator and characterizes the strength distribution of
the sp state $\alpha$ as a function of energy in the
$A-1$-particle system.
From this quantity one can therefore obtain
another key ingredient that gauges the
effect of correlations, namely the occupation number 
\begin{equation}
n(\alpha) = \int_{-\infty}^{\epsilon_F} d\omega\ S_h(\alpha , \omega )
= \bra{\Psi_0^A} a_\alpha^\dagger a_\alpha \ket{\Psi_0^A} .
\label{eq:occ}
\end{equation}
In the experimental analysis for the quasihole states
the quantum number $\alpha$ is related to the
Woods-Saxon potential required to both reproduce the correct energy of
the hole state as well as the shape of the corresponding
\eep\ cross section for the particular transition under consideration. 
The remaining parameter required to fit the data then 
becomes the spectroscopic factor associated with this transition.
In this analysis the reduction of the flux associated with the
scattering of the outgoing proton is incorporated by the use of empirical
optical potentials describing elastic proton-nucleus scattering data. 
Experiments on ${}^{208}{\rm Pb}$ result in a spectroscopic factor of 0.65
for the removal of the last $3s_{1/2}$ proton [2].
Additional information about the occupation number of this orbit can be 
obtained by analyzing elastic electron scattering cross sections of 
neighboring nuclei [20].
The occupation number for the $3s_{1/2}$ proton orbit obtained
from this analysis is about 10\% larger than the quasihole spectroscopic
factor [21,22,2].
A recent analysis of the \eep\ reaction on ${}^{208}{\rm Pb}$
in a wide range of missing energies and for missing momenta below 270 MeV/c
yields information on the occupation numbers of all the 
more deeply-bound proton orbitals.
The data suggest that all these deeply-bound orbits are depleted by the same
amount of about 15\% [23,24].

The properties of the 
experimental strength distributions can be understood on the basis of
the coupling between single-hole states and 2h1p states.
This implies that a proper inclusion of this coupling in the low-energy
domain is required in theoretical calculations that aim at reproducing the
experimental distribution of the strength.
Such calculations have been successfully performed for medium-heavy 
nuclei [25-27].
Indeed, calculations for the strength distribution for the removal of protons
from ${}^{48}{\rm Ca}$ demonstrate that an excellent qualitative agreement
with the experimental results is obtained when the coupling of the single-hole
states to low-lying collective states is taken into account [27].
This coupling is taken into account by calculating the microscopic RPA
phonons and then constructing the corresponding self-energy.
The solution of the Dyson equation then provides the theoretical
strength distribution [6].
By adding the additional depletion due to SRC, a
quantitative agreement can be obtained although no explicit calculation
for these nuclei including both effects has been performed to date.
The corresponding occupation numbers calculated for this nucleus
also indicate that the influence of collective low-lying states, associated
with long-range correlations, on the occupation numbers is confined to
sp states in the immediate vicinity of the
Fermi level [27].
The experimental information on occupation numbers in ${}^{208}$Pb [23,24]
suggests therefore that the observed depletion for deeply bound states
is essentially only due to SRC as will be further discussed in Sect.~3.
The description of the spectroscopic strength in ${}^{16}$O is not as 
successful [28] on account of the complexity of the low-energy
structure of this nucleus.
Although the results of Ref.~[28] demonstrate the importance of long-range
correlations for this nucleus, the final results for the $p$-quasihole strength
is still 0.2 above the data [16] which yield about 0.6 for the corresponding 
spectroscopic factors.
It should be noted that the inclusion of SRC only yields a 10\% reduction of
the strength [9,14], while center-of-mass corrections raise these spectroscopic
factors by about 7\% [29].
Attempts to describe the proper inclusion of microscopic particle-particle
and particle-hole phonons in a Faddeev approach for this nucleus are currently
in progress [30,31].
The Faddeev approach is necessary since the naive idea of adding the 
contribution to the self-energy of particle-particle and particle-hole phonons
while subtracting the common second-order term fails.
This failure is particularly salient in finite systems since near the poles
of the second-order self-energy no proper solution of the Dyson equation can
be obtained [30].

For a quantitative understanding it is also
necessary to account for the appearance of sp strength
at high momenta as a direct reflection of the influence of SRC.
These high-momentum nucleons make up an important part of
the missing strength that has been documented in \eep\ experiments.
Results for ${}^{16}{\rm O}$ [9,14] corroborate the expected
occupation of high-momenta but put their presence at high missing
energy.
This can be understood in terms of the admixture of a high-momentum nucleon
requiring 2h1p states which must accomodate this momentum maintaining momentum
conservation.
Since two-hole states combine to small total pair momenta, one necessarily
needs a high-momentum nucleon (of about equal and opposite value to the
component to be admixed) with corresponding high excitation energy.
As a result, one expects to find high-momentum components predominantly
at high missing energy.
Recent experiments at JLab are aimed at a quantitative assessment of the
strength distribution of these high-momentum nucleons [32].
A successful determination of this experimental strength would
finally complete the search for all the protons in the nucleus.
Sofar, only little more than 80\% of them have been identified for
${}^{208}$Pb [23].

\section{Considerations regarding Saturation Properties of Nuclear Matter} 
We will now focus on the consequences of the results discussed in the 
previous section. We start by arguing
that the empirical saturation density
of nuclear matter is dominated by SRC. 
As discussed earlier, a recent analysis of the \eep\ reaction on
${}^{208}{\rm Pb}$ up to 100 MeV missing energy and 270 MeV/c missing momenta
indicates that all deeply bound orbits are depleted by the same
amount of about 15\% [23,24].
This global depletion of the sp strength in about the same amount
for all states as observed for ${}^{208}{\rm Pb}$
was anticipated [5,33] on the basis of the experience that has been
obtained with calculating occupation numbers in nuclear matter with the
inclusion of SRC [34,35].
Such calculations suggest that about 15\% of the sp strength
in heavy nuclei is removed from the Fermi sea leading to the occupation of
high-momentum states. This global depletion of mean-field orbitals
can be interpreted as a clear signature of
the influence of SRC.
In turn, these results reflect on one of the key quantities determining
nuclear saturation empirically.
Elastic electron scattering from ${}^{208}{\rm Pb}$ [36]
clearly pinpoints the value of the central charge density in this nucleus.
By multiplying this number by $A/Z$ one obtains the relevant central density
of heavy nuclei, corresponding to 0.16 nucleons/${\rm fm}^3$ or $k_F = 1.33~ 
{\rm fm}^{-1}$.
Since the presence of nucleons at the center of a nucleus is confined
to $s$ nucleons, and their depletion is dominated by SRC,
one may conclude that the actual value of the saturation density of
nuclear matter must also be closely linked to the effects
of SRC.
While this argument is particularly appropriate for the deeply bound
$1s_{1/2}$ and $2s_{1/2}$ protons, it continues to hold for the $3s_{1/2}$ 
protons which are depleted predominantly by short-range effects (up to 15\%)
and by at most 10\% due to long-range correlations [2,21,22].

The binding energy of nuclei or nuclear matter usually
includes only mean-field contributions to the kinetic energy when 
the calculations
are based on perturbative schemes like the hole-line expansion [37].
With the presence of high-momentum components in the ground state it becomes
relevant to ask what the real kinetic and potential energy of the system
look like in terms of the sp strength distributions.
This theoretical result [18,19] has the general form
\begin{equation}
E_0^A = \bra{\Psi_0^A} \hat{H} \ket{\Psi_0^A}
= \half \sum_{\alpha\beta} \bra{\alpha} T \ket{\beta} n_{\alpha\beta}
+\half \sum_\alpha \int_{-\infty}^{\epsilon_F}d\omega\ \omega 
S_h(\alpha, \omega)
\label{eq:be}
\end{equation}
in the case when only two-body interactions are involved.
In this equation, $n_{\alpha\beta}$ is the one-body density matrix element
which can be directly obtained from the sp propagator.
Obvious simplifications occur in this result for the case of nuclear matter 
due to momentum conservation.
A delicate balance exists between the repulsive kinetic-energy term
and the attractive contribution of the second term in Eq.~(\ref{eq:be})
which samples the sp strength weighted by the energy
$\omega$.
When realistic spectral distributions are used to calculate
these quantities in finite nuclei unexpected results emerge [14].
Such calculations for ${}^{16}{\rm O}$ indicate that the
contribution of the
quasihole states to Eq.~(\ref{eq:be}),
comprises only 37\% of the total energy
leaving 63\% for the continuum terms that represent the spectral strength
associated with the coupling to low-energy 2h1p states.
The latter contributions exhibit the presence of high-momentum
components in the nuclear ground state.
Although these high momenta 
account for only 10\% of the particles in ${}^{16}{\rm O}$, their
contribution to the energy is extremely important.
These results demonstrate the importance of treating
the dressing of nucleons in finite nuclei in determining the binding
energy per particle.
It is therefore reasonable to conclude that a careful study of 
SRC including the full fragmentation of the sp
strength is necessary for the calculation of the energy per particle
in finite nuclei.
Such considerations for nuclear matter have been available for some time
as well [38].
Including fragmentation of the sp strength has 
the additional advantage that agreement with data from the
\eep\ reaction [32] can be used to gauge the quality of the
theoretical description in determining the energy per particle.
This argument can be turned inside out by noting that an exact representation
of the spectroscopic strength must lead to the correct energy
per particle according to Eq.~(\ref{eq:be}) in the case of the
dominance of two-body interactions. Clearly this perspective can only become
complete upon the successful analysis of high-momentum components in the
\eep\ reaction [32].

Returning to the saturation problem in nuclear matter, it is
important to comment on the recent success of the Catania group in determining
the nuclear saturation curve including three hole-line 
contributions [37].
These calculations demonstrate that a good agreement is obtained at the
three hole-line level between 
calculations that start from different prescriptions
for the auxiliary potential.
Since the contribution of the three hole-line terms are 
significant but indicate reasonable convergence properties compared to
the two hole-line contribution, one may assume that
these results provide an accurate representation of the energy per particle
as a function of density for the case of only nonrelativistic nucleons.
The saturation density obtained in this recent work corresponds to $k_F =
1.565~{\rm fm}^{-1}$ with a binding energy of -16.18 MeV.
The conclusion appears to be appropriate that additional physics
in the form of three-body forces or the inclusion of relativistic effects
is necessary to repair this obvious discrepancy with the empirical saturation
properties.

Before agreeing with this conclusion it is useful to remember that three
hole-line contributions include a third-order ring diagram.
The agreement of three hole-line calculations with advanced variational
calculations [39] further emphasizes the notion that important
aspects of long-range correlations are included in both these calculations.
This conclusion can also be based on the observation that hypernetted chain
calculations effectively include ring-diagram contributions
to the energy per particle although averaged over the Fermi sea [40]. 
The effect of these long-range correlations on nuclear saturation properties
is not small and can be illustrated by quoting explicit results for three-
and four-body ring diagrams [41].
These results for the Reid potential [42],
including only nucleons, demonstrate that these ring-diagram terms
are dominated by attractive
contributions involving pion quantum numbers propagating
around the rings.
Furthermore, these contributions increase in importance with increasing
density.
Including the possibility of the coupling of these pionic excitation modes
to $\Delta$-hole states in 
these ring diagrams leads to an additional large increase in the binding
with increasing density [41].
Alternatively, these terms involving $\Delta$-isobars
can also be considered as contributions due to three- and four-body
forces in the space of only nucleons.
The importance of these long-range contributions to the binding energy
is of course related to the possible appearance of pion condensation
at higher nuclear density.
These long-range pion-exchange dominated contributions to the
binding energy appear because of conservation of momentum in nuclear
matter. For a given momentum $q$ carried by a pion around a ring diagram,
one is able to sample coherently the attractive interaction that exists
for values of $q$ above 0.7 ${\rm fm}^{-1}$.
All ring diagrams contribute coherently when the interaction is attractive
and one may therefore obtain huge contributions at higher densities
which reflect the importance of this collective pion-propagation 
mode [43].

No such collective pion-degrees of freedom are actually observed
in finite nuclei.
A substantial part of the explanation of this fact is provided by the
observation that in finite nuclei both the attractive and repulsive parts of
the pion-exchange interaction are sampled before a build-up
of long-range correlations can be achieved.
Since these contributions very nearly cancel each other, which is
further facilitated by the increased relevance of exchange terms [44],
one does not see any marked effect on pion-like excited states in nuclei
associated with long-range pion degrees of freedom
even when $\Delta$-hole states are included [45].
It seems therefore reasonable to call into question the relevance
of these coherent long-range pion-exchange contributions
to the binding energy per particle in nuclear matter.
Since the actual saturation properties of nuclei appear to be
dominated by SRC, as discussed above,
a critical test of this idea may be to calculate nuclear saturation
properties focusing solely on the contribution of SRC.
The recent experimental results discussed above demand furthermore
that the dressing of nucleons in nuclear matter is taken into account
in order to be consistent with the extensive collection of experimental
data from the \eep\ reaction
that have become available in recent years.
The self-consistent calculation of nucleon spectral functions
obtained from the contribution to the nucleon self-energy
of ladder diagrams which include
the propagation of these dressed particles, fulfills this requirement.
Some details of this scheme will be discussed in the next section together
with the first results [17,11].

\section{Self-consistently Dressed Nucleons in Nuclear Matter} 
It is straightforward to write down the equation that involves
the calculation of the effective interaction in nuclear matter
obtained from the sum of
all ladder diagrams while propagating fully dressed particles.
This result is given in a partial wave representation by the following
equation
\begin{eqnarray}
& {} &\bra{k}\Gamma_{LL'}^{JST}(K,\Omega)\ket{k'} = 
\bra{k}V_{LL'}^{JST}(K,\Omega)\ket{k'} \nonumber \\
& + & \sum_{L''} \int_0^\infty dq\ q^2\
\bra{k}V_{LL''}^{JST}(K,\Omega)\ket{q} g_f^{II}(q;K,\Omega)
\bra{q}\Gamma_{LL''}^{JST}(K,\Omega)\ket{k'} ,
\label{eq:lad}
\end{eqnarray}
where $k,k',$ and $q$ denote relative and $K$ the total momentum
involved in the interaction process.
Discrete quantum numbers correspond to total
spin, $S$, orbital angular momentum, $L,L',L''$, and
the conserved
total angular momentum and isospin, $J$ and $T$, respectively.
The energy $\Omega$ and the total momentum $K$ are conserved and act
as parameters that characterize the effective two-body interaction in the
medium.
The critical ingredient in Eq.~(\ref{eq:lad}) is the noninteracting
propagator $g_f^{II}$ which describes the propagation of the particles
in the medium from interaction to interaction.
For fully dressed particles this propagator is given by
\begin{eqnarray}
g_f^{II}(k_1,k_2;\Omega) & = &
\int_{\epsilon_F}^\infty d\omega_1\ \int_{\epsilon_F}^\infty d\omega_2\
\frac{S_p(k_1,\omega_1) S_p(k_2,\omega_2)}
{\Omega - \omega_1 -\omega_2 +i\eta} \nonumber \\
& - &
\int_{-\infty}^{\epsilon_F} d\omega_1\ \int_{-\infty}^{\epsilon_F} d\omega_2\
\frac{S_h(k_1,\omega_1) S_h(k_2,\omega_2)}
{\Omega - \omega_1 -\omega_2 -i\eta} ,
\label{eq:gtwof}
\end{eqnarray}
where individual momenta $k_1$ and $k_2$ have been used instead
of total and relative momenta as in Eq.~(\ref{eq:lad}).
The dressing of the particles is expressed in the use of particle and
hole spectral functions, $S_p$ and $S_h$, respectively.
The particle spectral function, $S_p$, is defined as a particle
addition probability density in a similar way
as the hole spectral function in Eq.~(\ref{eq:spefu}) for removal.
These spectral functions take into account that the particles propagate
with respect to the correlated ground state incorporating the
presence of high-momentum components in the ground state.
This treatment therefore provides the correlated version of the Pauli
principle and leads to substantial modification with respect to
the Pauli principle effects related to the free Fermi gas.
This fact suggests that this correlated version may also provide a
reasonable description at higher densities since the propagation of
particles is considered with respect to the correlated ground state.
The propagator corresponding to the Pauli principle of the free Fermi gas
is obtained from Eq.~(\ref{eq:gtwof}) by
replacing the spectral functions by strength distributions characterized
by $\delta$-functions as follows
\begin{eqnarray}
S_p(k,\omega) & = & \theta(k-k_F) \delta(\omega-\epsilon(k)) \nonumber \\
S_h(k,\omega) & = & \theta(k_F-k) \delta(\omega-\epsilon(k)) .
\label{eq:mf}
\end{eqnarray}
This leads to the so-called Galitski-Feynman propagator including hole-hole
as well as particle-particle propagation of particles characterized
by sp energies $\epsilon(k)$.
Discarding the hole-hole propagation then yields the 
Brueckner ladder diagrams with the usual Pauli operator for the free
Fermi gas.
The effective interaction obtained by solving Eq.~(\ref{eq:lad}) using
dressed propagators can be used to construct the self-energy
of the particle. With this self-energy the Dyson equation can be solved
to generate a new incarnation of the dressed propagator.
The process can then be continued by constructing anew the dressed
but noninteracting two-particle propagator according to Eq.~(\ref{eq:gtwof}).
At this stage, one can return to the ladder equation and so on until
self-consistency is achieved for the complete Green's function
which is then legitimately called a self-consistent Green's function.

While this scheme is easy to present in equations and words, it is quite
another matter to implement it.
The recent accomplishment of implementing this self-consistency
scheme [17] builds upon earlier approximate implementations.
The first nuclear-matter spectral functions were obtained for a semirealistic
interaction by employing mean-field propagators in the ladder
equation [46].
Spectral functions for the Reid interaction were obtained by still employing
mean-field propagators in the ladder equation but with the introduction of
a self-consistent gap in the sp spectrum to take into account
the pairing instabilities obtained for a realistic interaction [38,47].
The first solution of the effective interaction using dressed propagators
was obtained by employing a parametrization of the spectral 
functions [48,49].
The calculations employing dressed propagators in determining
the effective interaction
demonstrate that at normal density one no longer runs into
pairing instabilities on account of the reduced density of states associated
with the reduction of the strength of the quasiparticle pole, $z_{k_F}$,
from 1 in the Fermi gas to 0.7 in the case of dressed propagators.
For two-particle propagation this leads to a reduction factor of
$z_{k_F}^2$ corresponding to about 0.5 that is strong enough to push
even the pairing instability in the ${}^3S_1$-${}^3D_1$ channel
to lower densities [49].
The consequences for the scattering process of
interacting particles in nuclear matter characterized
by phase shifts and cross sections are also substantial and lead to a
reduction of the cross section in a wide range of energies [49].

The current implementation of the self-consistent scheme for the
propagator across the summation of all ladder diagrams
includes a parametrization of the imaginary part of the nucleon self-energy.
Employing a representation in terms of two gaussians above and two below
the Fermi energy, it is possible to accurately represent the
nucleon self-energy as generated by the contribution of relative $S$-waves
(and including the tensor coupling to the ${}^3D_1$ channel) [17].
Self-consistency at a density corresponding to $k_F = 1.36\ {\rm fm}^{-1}$
is achieved in about ten iteration steps, each involving a considerable
amount of computer time [17].
A discrete version of this scheme is being implemented successfully by the
Gent group [50,51].
It is important to reiterate that such
schemes isolate the contribution of SRC 
to the energy per particle which is obtained from Eq.~(\ref{eq:be}).
\begin{figure}[htb]
  \begin{center}
    \includegraphics[height=0.40\textheight]{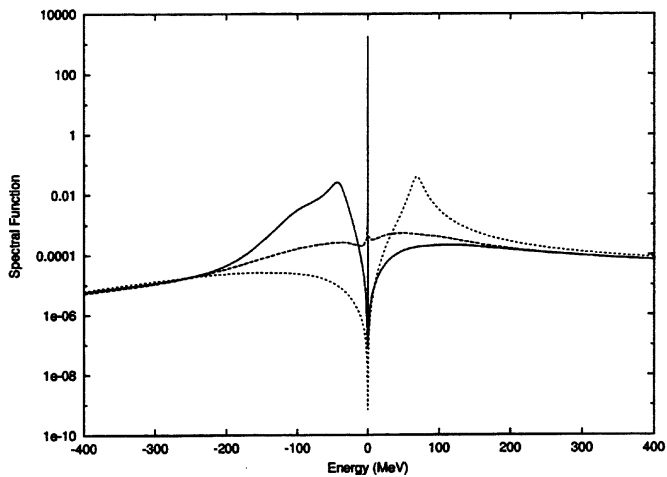}
    \caption{\label{fig:spfu}
      Self-consistent spectral functions at $k_F = 1.36$ ${\rm fm}^{-1}$.
      Single-particle momenta corresponding to $k = 0$ (solid), $k_F$ (dashed),
      and 2.1 ${\rm fm}^{-1}$ (dotted) are shown.
      }
    \end{center}
\end{figure}
An important result pertaining to this ``second generation'' spectral functions
is shown in Fig.~\ref{fig:spfu} related to the emergence of a common tail
at large negative energy for different momenta. Such a common 
tail was previously obtained at high energy [35] in the particle domain
as a signature of SRC.
This common tail appears to play a significant role in generating some
additional binding energy at lower densities compared to conventional
Brueckner-type calculations.
At present, results for two densities corresponding to $k_F = 1.36$ and 1.45
${\rm fm}^{-1}$ have been obtained.
Self-consistency is achieved for the contribution of the ${}^1S_0$ and 
${}^3S_1$-${}^3D_1$ channels to the self-energy. The other partial wave 
contributions have been added separately.
In practice, higher partial waves are always included in the correlated 
Hartree-Fock contribution. We have obtained additional contributions for
$L = 2$ and 3 from solutions of the dressed ladder equation after obtaining
self-consistency with the dominant $S$ waves.
The corresponding results for the
binding energy have been obtained by averaging the parametrizations of the
corresponding self-energies with and without these higher-order terms for 
$L = 2$ and 3 partial waves. The difference between these two results
then provides us with a conservative estimate
of the lack of self-consistency including these terms in higher partial waves.
\begin{figure}[htb]
  \begin{center}
    \includegraphics[height=0.30\textheight]{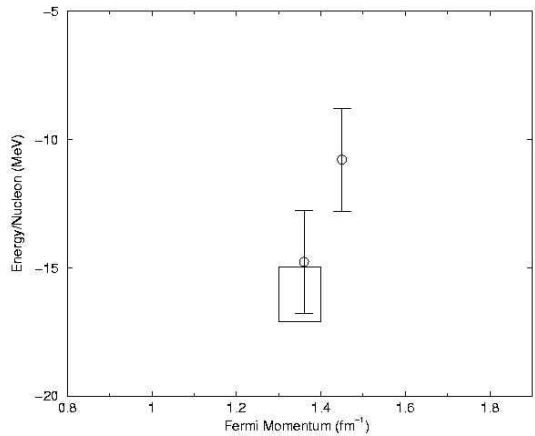}
    \caption{\label{fig:bea2}
      The energy per particle calculated at two densities. The saturation 
      density for this self-consistent Green's function calculation with the
      Reid potential is possibly in agreement with the empirical result.
      }
    \end{center}
\end{figure}
This error estimate is included in Fig.~\ref{fig:bea2} for the energy per
particle calculated from the energy (Koltun) sum rule in Eq.~(\ref{eq:be}). 
These results suggest that it is possible to obtain reasonable saturation
properties for nuclear matter provided one only includes SRC in the 
determination of the equation of state.
Based on the arguments presented in the previous section, one should not
be too surprised with this result.
Clearly, the assertion that long-range pion-exchange contributions
to the energy per particle need not be considered in explaining nuclear
saturation properties, needs to be further investigated.
In practice, this means that one needs to establish whether pion-exchange
in heavy nuclei already mimics the corresponding process in nuclear matter.
If this does not turn out to be the case, the arguments for considering
the nuclear-matter saturation problem only on the basis of the contribution
of SRC will be strenghtened considerably.
Furthermore, one would then also expect that the contribution of three-body
forces [52] to the binding energy per particle in finite nuclei continues to be
slightly attractive when particle number is increased substantially beyond 10
[53].
This point and the previous discussion also suggest that there
would be no further need for the ad-hoc repulsion added to three-body forces
used to fit nuclear-matter saturation properties [54]. 

\section{Conclusions}
One of the critical experimental ingredients in clarifying the nature
of nuclear correlations has only become available over the last decade
and a half. It is therefore not surprising that all schemes that
have been developed to calculate nuclear-matter saturation properties
are not based on the insights that these experiments provide.
One of the aims of the present paper is to remedy this situation.
To this end we have started with
a review of experimental data obtained from the \eep\ reaction
and corresponding theoretical results,
that exhibit clear evidence that nucleons in nuclei exhibit strong correlation
effects.
Based on these considerations and the success of the theoretical calculations
to account for the qualitative features of the sp
strength distributions, it is suggested that the dressing of nucleons
must be taken into account in calculations of the energy per particle.
By identifying the dominant contribution of SRC 
to the empirical saturation density, it is argued that these correlations
need to be emphasized in the study of nuclear matter.
It is also argued that inclusion of long-range correlations, especially those
involving pion propagation, leads to an unavoidable increase in the
theoretical saturation density.
Since this collectivity in the pion channel is not observed in nuclei,
it is proposed that the corresponding correlations in nuclear matter
are not relevant for the study of nuclear saturation and should therefore
be excluded from consideration.
A scheme which fulfills this requirement and includes the propagation
of dressed particles, as required by experiment, is outlined.
Successful implementation of this scheme has recently been
demonstrated [17,51]. First results demonstrate that these new
calculations lead to substantially lower saturation densities than 
have been obtained in the past.
The introduction of a ``nuclear-matter problem'' which focuses solely
on the contribution of SRC may therefore lead to new insight into
the long-standing problem of nuclear saturation.

\section*{Acknowledgments}
This work was supported by the U. S. National Science Foundation under Grant
No. PHY-9900713.

\end{document}